\documentclass[prl,twocolumn,aps,showpacs]{revtex4} % showkeys

\usepackage{graphicx}% Include figure files
\usepackage{dcolumn}% Align table columns on decimal point
\usepackage{bm}% bold math
\usepackage{longtable}

\begin{document}

\title{Quantifying age- and gender-related diabetes comorbidity risks using nation-wide big claims data}

\author{Peter Klimek$^1$}
\author{Alexandra Kautzky-Willer$^2$}
\author{Anna Chmiel$^1$}
\author{Irmgard Schiller-Fr\"uhwirth$^3$}
\author{Stefan Thurner$^{1,4,5}$}
\email{stefan.thurner@meduniwien.ac.at}
\affiliation{$^1$Section for Science of Complex Systems, Medical University of Vienna, Spitalgasse 23, A-1090, Austria.
$^2$Gender Medicine Unit, Endocrinology \& Metabolism, Dept. of Internal Medicine III, Medical University of Vienna, Spitalgasse 23, A-1090, Austria.
$^3$Main Association of Austrian Social Security Institutions, Kundmanngasse 21, A-1031, Austria. 
$^4$Santa Fe Institute, 1399 Hyde Park Road, Santa Fe, NM 87501, USA.
$^5$IIASA, Schlossplatz 1, A-2361 Laxenburg, Austria.
}
\email{stefan.thurner@meduniwien.ac.at}
\date{\today}

\begin{abstract}
Currently emerging 'big data' techniques are reshaping medical science into a data science. Medical claims data allow assessing an entire nation's health state in a quantitative way, in particular with regard to the occurrences and consequences of chronic and pandemic diseases like diabetes.
We develop a quantitative, statistical approach to test for associations between the incidence of type 1 or type 2 diabetes and any possible other disease as provided by the ICD10 diagnosis codes using a complete set of Austrian inpatient data. With a new co-occurrence analysis the relative risks for each possible comorbid disease are studied as a function of patient age and gender, a temporal analysis investigates whether the onset of diabetes typically precedes or follows the onset of the other disease. The samples is always of maximal size, i.e. contains all patients with that comorbidity within the country. The present study is an equivalent of almost 40,000 studies, all with maximum (complete) patient number available in the country. 
Out of more than thousand possible associations, 123 comorbid diseases for type 1 or type 2 diabetes are identified at high significance levels.
Well known diabetic comorbidities are recovered, such as retinopathies, hypertension, overweight, chronic kidney diseases, etc. This validates the method. Additionally, a number of comorbidities are identified which have only been recognized to a lesser extent, for example epilepsy, sepsis, or several mental disorders. The temporal evolution, age, and gender-dependence of these comorbidities are discussed. The new statistical-network methodology developed here can be readily applied to other chronic diseases.

\end{abstract}

\maketitle

\section{Background}

The emerging availability of 'big data' transforms medical sciences into a data science. Nation-wide collections of physician and hospital claims data allow exploring the health state of an entire country's population with unprecedented precision and scale \cite{ref1}. The enormous potential of such claims data has been shown by developing or improving data-driven comorbidity indices to predict mortality rates \cite{ref2} , or by studying healthcare utilization and outcome measures of specific patient cohorts \cite{ref3}. In this work we show how to use a nation-wide claims dataset to generate and cross-validate hypothesis on each possible comorbidity association for type 1 (DM1) and 2 diabetes (DM2), and how they depend on patient age, gender, and temporal order of the disorders' onsets. 

Diabetes is a global pandemic disease. The world-wide number of adult diabetes patients doubled over the last three decades to approximately 350 million as of 2010, and is expected to double again until 2030 as a result of population ageing and a shift to western lifestyle patterns in developing countries \cite{ref4}. Diabetes comprises a heterogeneous group of disorders with the most prominent types being DM1 and DM2. These disorders have different pathophysiologies and phenotypes; the exact underlying mechanisms, their interplay finally leading to manifestation, progressions of the diseases, and their complications are still unclear. Diabetes is associated with a large number of comorbid diseases, including but not limited to vascular complications \cite{ref5}, renal failures \cite{ref5}, neuropathy \cite{ref5}, heart diseases \cite{ref6,ref7}, cognitive disorders \cite{ref8,ref9}, retinopathy \cite{ref10}, and hypertension \cite{ref11}. Each of these comorbidities opens up a unique direction of research. Following the methodological approach developed in this work, thousands of such associations can be investigated in parallel. Besides studying the individual diabetic comorbidities, and how they depend on patient age and gender, this allows to compare the strength of these associations among each other and rank them according to their incidences. The methodology developed here can be readily applied to obtain complete comorbidity profiles for other disorders than diabetes.

\section{Methods}

{\bf Data.} A database of the Main Association of Austrian Social Security Institutions containing pseudonymised claims data of all persons receiving out- and inpatient care in Austria between January 1st, 2006 and December 31st, 2007 is used \cite{ref12}. The data gives a comprehensive, nation-wide picture of the medical condition of most of the approximately 8.3 million Austrians. The patient collective was formed by extracting all persons receiving inpatient care in 2006 or 2007. We identified patients being diagnosed with DM1 or DM2 (ICD10 codes E10 and E11). Patients who died in 2006 or 2007 were removed. In this way 16,667 DM1 patients (8,355 males and 8,312 females) and 105,904 with DM2 (50,596 males and 55,308 females) were selected. The total sample of inpatients consists of 1,862,258 patients (1,064,952 females and 797,306 males). From these patients we know their year of birth, sex, ATC codes of all their prescriptions, and the ICD codes of all their diagnoses (main- and side-diagnoses). 

{\bf Co-occurrence analysis.} For the occurrences of each diagnosis x (ICD10, three-digit-level) a patient-age-resolved cross tabulation with the occurrences of DM1 and DM2 is performed. Symptoms, injuries, pregnancies, and external causes and factors of morbidity were excluded. The patients are grouped according to their age in five-year intervals. For each diagnosis and age interval a contingency table is built. If each entry in the table is greater than 10, relative risks $RR_{1(2)}(x,t)$ are computed, a chi-squared test is performed and $p$-values are calculated for rejecting the null hypothesis that co-occurrence of the diagnoses with DM1 or DM2 is independent. The Benjamini-Hochberg procedure \cite{ref13} is applied to control for the false discovery rate $\alpha$. If there are less than ten co-occurrences or the results are not significant, the relative risk is set to one.

{\bf Gender ratio.} The gender ratio $GR(x,t)$ is related to the quotient of the percentage of female and male diabetes patients in age group $t$ who also have diagnoses $x$, see supplementary information.

{\bf Lead/lag indicator.} The lead/lag indicators assess whether patients with diagnoses $d_i$ are more likely to be later in their life diagnosed with another disease $x$, the lead indicator $I_{lead} (d_i,x)$, or whether it is more likely that people having diagnoses $x$ will be diagnosed with diabetes, the lag indicator $I_{lag}(d_i,x)$, see supplementary information.

\section{Results}

\begin{figure*}[tbp]
 \begin{center}
 \includegraphics[width=165mm]{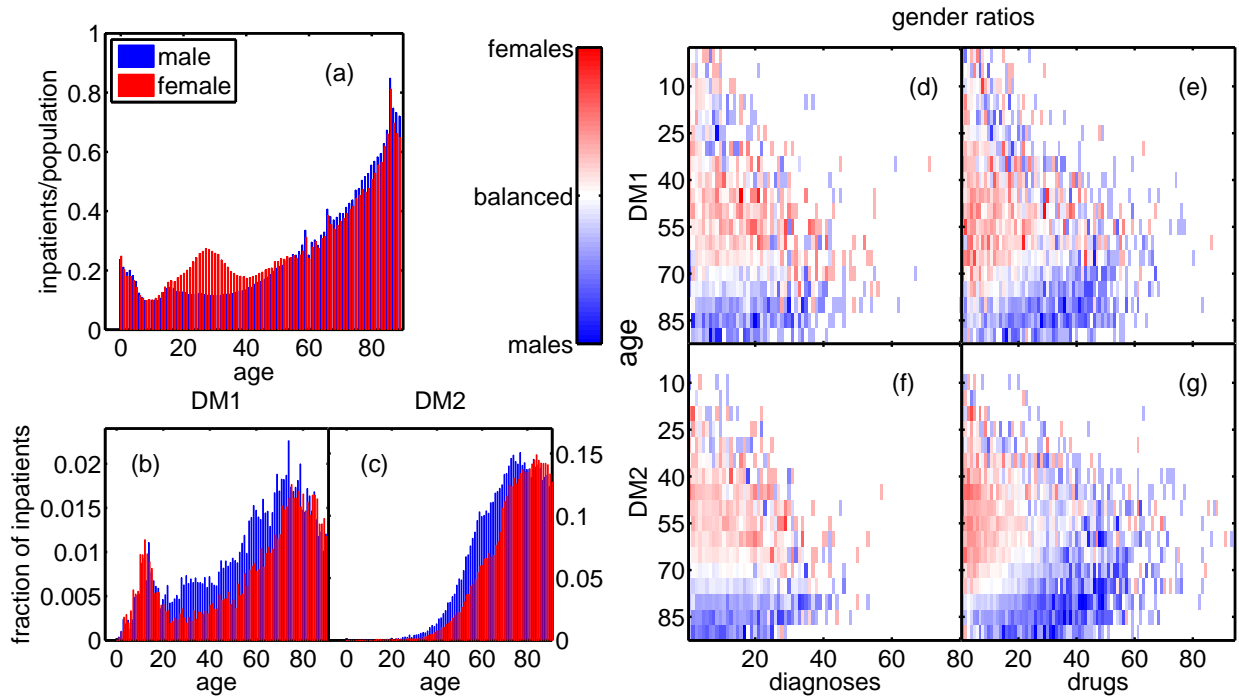}
  \end{center}
 \caption{The fraction of inpatients in the entire population as a function of age is similar in males and females, except for an excess of females at the age around 30, most likely related to giving birth (a). After a peak in early childhood, the fraction of inpatients increases to levels of above 80\% in older age. The bulk of male (female) (b) DM1 and (c) DM2 patients is aged around 60 (70), for DM1 patients there is a second peak around age 10. The gender ratio $GR(y,t)$ is shown for DM1 (d,e) and DM2 (f,g) patients and the number of their diagnoses (d,f) and their prescriptions (e,g). For patients younger than 60, with a comparably high number of comorbidities, female patients have less diagnoses but take more drugs than males.}
  \label{Fig1}
 \end{figure*}

Figure \ref{Fig1}(a) shows the fraction of male and female inpatients of the entire population as a function of age. The inpatient fractions are around 20\% for children under five, then drop to 10-15\% for ages around ten, and from then on rise to more than 80\% for 80 year-old patients, with an additional peak for females of age around 30, most likely due to child birth. Figure \ref{Fig1}(b) shows the fraction of male and female DM1 inpatients as a function of age. The distributions are bimodal with one peak around the typical onset-age of ten for both male and females, and a second peak for ages 60 (70) for males (females). Figure \ref{Fig1}(c) shows the fractions of inpatients diagnosed with DM2 as a function of age, with comparably few patients below age thirty, and the bulk of male (female) patients concentrated around age 60 (70).

Figure \ref{Fig1} shows $GR(y,t)$ for DM1 patients and their number of diagnoses (d) and received drugs (e), (f) and (g) show the same for DM2 patients. Up to an age of 60 there is an excess of male patients, for older patients there is an excess of females. For drugs there is a male excess only for age up to 60 and for less than 10-20 drugs. For older age and a larger number of drugs there is an excess of female patients. Females below age 60 have thus fewer diagnoses than males, but especially those with a large number of diagnoses have more prescriptions than males. After age 60, females outweigh males in both diagnoses and prescriptions.

\begin{figure*}[tbp]
 \begin{center}
 \includegraphics[width=165mm]{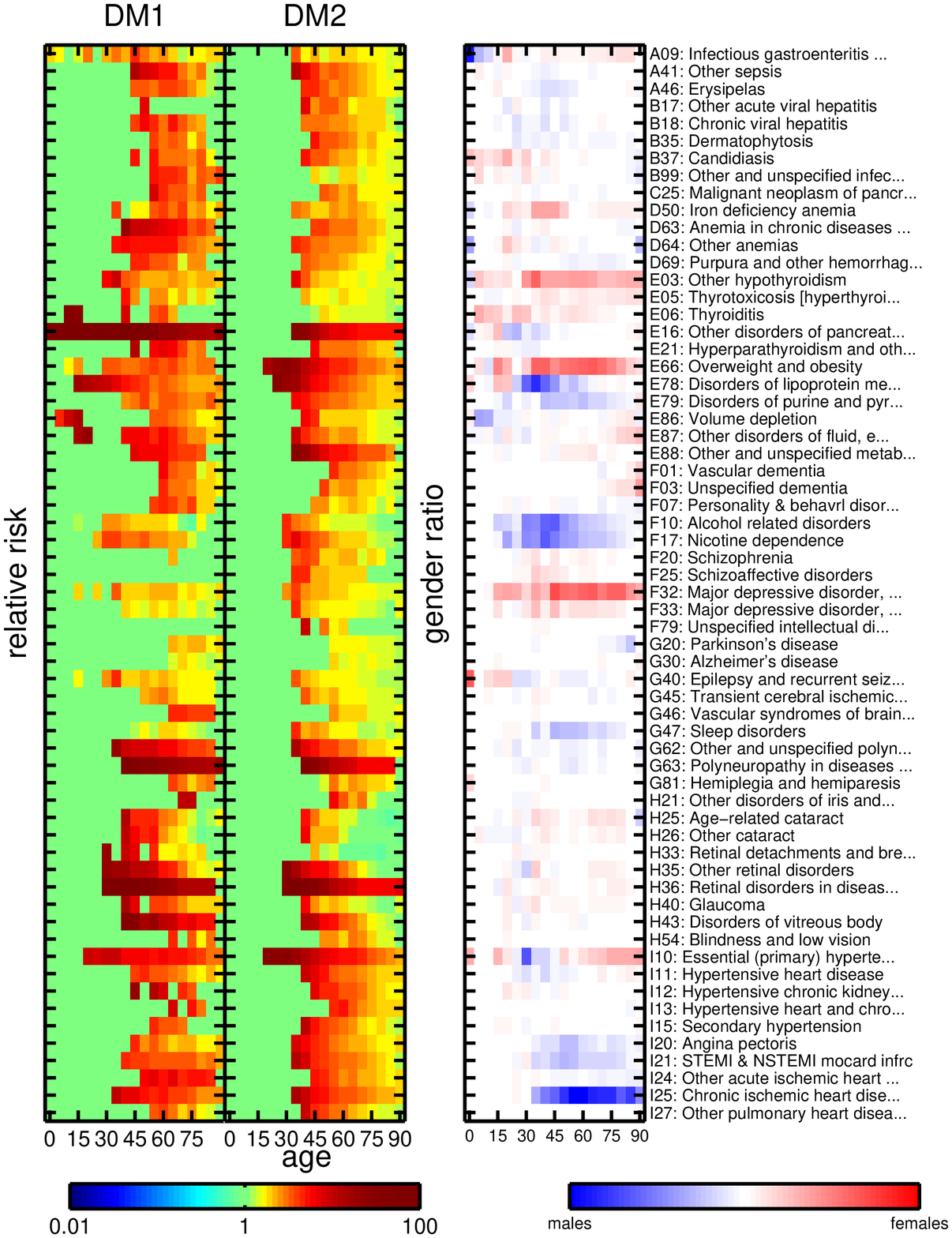}
  \end{center}
 \caption{Relative risks for DM1 (left column) and DM2 (middle column) patients, and gender ratios (right column) for core comorbidities using a false discovery rate of $\alpha<0.01$ and an ICD code from the range A01-I27. Color encodes the values of the risks and gender ratios.}
  \label{Fig2}
 \end{figure*}

\begin{figure*}[tbp]
 \begin{center}
 \includegraphics[width=165mm]{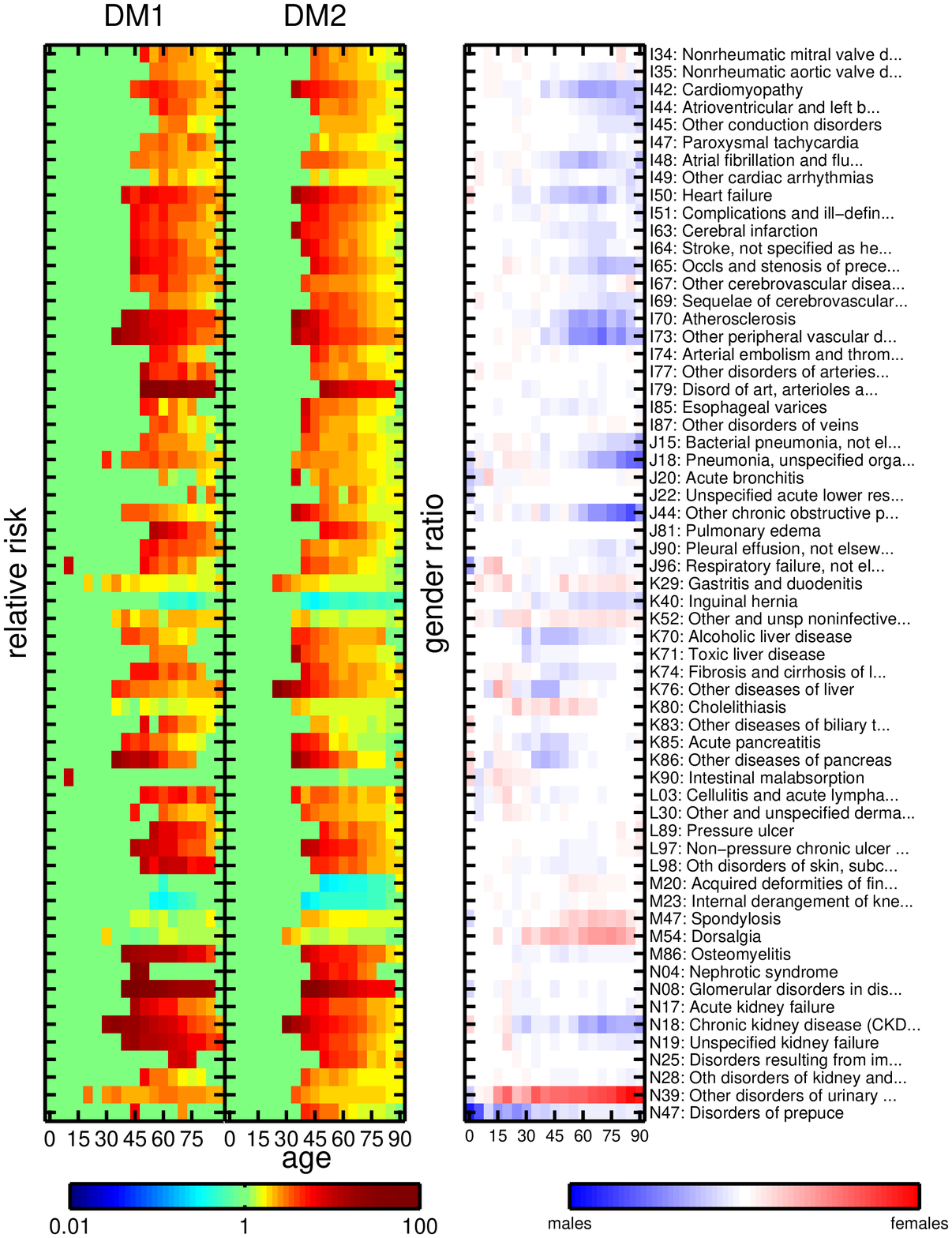}
  \end{center}
 \caption{Relative risks for DM1 (left column) and DM2 (middle column) patients, and gender ratios (right column) for core comorbidities using a false discovery rate of $\alpha<0.01$ and an ICD code from the range I34-N99. Color encodes the values of the risks and gender ratios.}
  \label{Fig3}
 \end{figure*}

{\bf Co-occurrence analysis.} Each diagnosis where the null hypothesis of statistical independence with either DM1 or DM2 can be rejected with a false discovery rate of $\alpha<0.01$ in at least one of the age groups is identified as a comorbidity. The results are summarized in Figure \ref{Fig2} and Figure \ref{Fig3}, with the left columns showing the DM1 relative risk $RR_1 (x,t)$, the middle columns the DM2 relative risk $RR_2 (x,t)$, and the right columns the gender ratio $GR(x,t)$. The comorbidities are also listed in the supplement, Table S1, along with relative risks, $p$-values, and patient ages for the age group with the smallest $p$-values for DM1 and DM2, respectively. In the following we refer to these values whenever referring to the relative risks of a diagnosis with a 95\% confidence interval (CI).

{\bf Lead/lag analysis.} To identify lead/lag behavior $I_{lead}(d_i,x)$ and $I_{lag}(d_i,x)$ are computed for male and female DM1 and DM2 patients. A leading or lagging relationship is identified if the null hypothesis that the observed indicator values can be obtained from randomized surrogate data can be rejected with a $p$-value of $p<0.05$, see the supplement for details on the randomization procedure. Table \ref{leadlag} shows diagnoses which have been identified as either leading or lagging for male or female DM1 or DM2 patients, as well as their corresponding $p$-values.

\begin{table*}[tbp]
\caption{Diagnoses are shown which have been identified in the lead/lag analysis. For each diagnoses the order (if diabetes leads or lags), gender ('F' for females, 'M' for males) and diabetes type (I or II) where the relationship was detected are listed.}
\begin{tabular}{l|l|l}
ICD (diagnosis) & diabetes type & gender \\
\hline
{\it \quad Diabetes leads (comes before the other disease)} & & \\
C25 Malignant neoplasm of pancreas &	II &	F\\
D63 Anaemia in chronic disease	&II	&F\\
D69 Purpura and other haemorrhagic conditions	&II	&F\\
E86 Volume depletion	&II	&F\\
F01 Vascular dementia	&II	&M + F\\
F07 Personality and behavioral disorders due to known physiological condition	&II	&F\\
F32 Depressive episode	&I	&M\\
G20 Parkinson's disease	&II	&M\\
G47 Sleep disorders	&II	&F\\
I25 Chronic ischaemic heart disease	&II	&M\\
I48 Atrial fibrillation and flutter	&II	&M + F\\
I50 Heart failure	&II	&F\\
J44 Other chronic pulmonary diseases	&II	&F\\
J90 Pleural effusion, not elsewhere classified	&II	&F\\
K52 Other and unspecified noninfective gastroenteritis and colitis	&II	&M\\
K83 Other diseases of intestine	&II	&M\\
L89 Decubitus ulcer	&II	&M + F\\
L97 Ulcer of lower limb n. e. c.	&II	&F\\
N18 Chronic renal failure	&II	&M + F \\
{\it \quad Diabetes lags (comes after the other disease)} & & \\
F10 Use of alcohol	&II	&M\\
F20 Schizophrenia	&II	&F\\
F25 Schizoaffective disorder	&II	&F\\
I20 Angina pectoris	&II	&F\\
I34 Nonrheumatic mitral valve disorders	&II	&M\\
I85 Esophageal varices	&II	&M\\
K70 Alcoholic liver disease	&II	&M\\
K86 Other diseases of pancreas	&II	&M\\
K90 Intestinal malabsorption	&I	&F\\
M20 Acquired deformities of fingers and toes	&II	&F
\end{tabular}
\label{leadlag}
\end{table*}

\section{Discussion}

{\bf Infections and sepsis.} Bacterial and viral infections can be found among the comorbidities (gastroenteritis, erysipelas, pneumonia, osteomyelitis, hepatitis \cite{ref14}, dermatophytosis \cite{ref15}, candidiasis). Most infections show an excess of male patients, exceptions are gastroenteritis and candidiasis which are dominated by female patients. It has been disputed whether diabetes increases susceptibility for sepsis or is potective \cite{ref16}. However diabetes is a common comorbity (incidence of 23\%) in septic patients \cite{ref17}. We find an excess of sepsis comorbidity which is strongest in male DM1 patients at the age around 50, with higher relative risks for DM1 (12, CI 8.2-18) than DM2 (2.7, CI 2.4-2.9). 

{\bf Neoplasms.} At present it is unclear if new onset diabetes is a specific marker of pancreatic cancer or if diabetic patients in general are at increased risk of pancreatic cancer with a pooled $RR$ of approximately 2 compared to non-diabetics in a meta-analysis \cite{ref18} with at least 1 year diabetes duration prior to diagnosis of pancreatic cancer \cite{ref19}. In our dataset there is an excess of male patients and higher relative risks for DM1 patients (8.6, CI 5.6-13) than DM2 patients (2.5, CI 2.1-2.8), potentially outlining greater impact of chronic hyperglycemia than of overweight-related parameters of the metabolic syndrome. It is identified as lagging behind the onset of DM2 in female patients, which might suggest that DM2 increases the risk of this cancer at least in females.

{\bf Anaemia.} Diabetes is a known risk factor for anaemia \cite{ref20}. Iron-deficiency and anaemia in chronic diseases show higher relative risks for DM1 (3.7, CI 3.0-4.6, and 6.3, CI 4.9-8.1, respectively) than DM2 (2.7, CI 2.4-2.9, and 2.8, CI 2.5-3.2) patients. There is an excess of female patients in the age range 20-40, thus being related to the reproductive phase.

{\bf Endocrine and metabolic disorders.} There are pronounced gender differences observable in endocrine and metabolic comorbidities \cite{ref5}. While patients with thyroiditis (mostly DM1), hypothyroidism, thyrotoxicosis, and obesity are predominantly female, disorders of the liporoteine, purine, and pyrimidine metabolism tend to be found in males. It is known that diabetic patients feature a 2-3 fold higher increased risk of disorders of the thyroid gland particularly those with autoimmune diabetes and that this association is strongly influenced by gender \cite{ref21}. For volume depletion and disorders of fluid, electrolyte and acid-base balance there appears to be an age switch, from an excess of male patients for younger ages to females in older age.

{\bf Mental and behavioral disorders.} Nicotine dependence and alcohol related disorders are comorbidities with excess-risks peaking at ages 30-45, dominated by male patients. For males the onset of DM2 lags behind the use of alcohol; alcoholic liver disease is also a lagging, male-dominated comorbidity. The relationship between alcohol consumption and DM2 has been shown to be dosage dependent. While moderate alcohol consumption was protective, dosages of more than 60g/day increased diabetes risk \cite{ref22}.
 
Depressions are dominated by female patients \cite{ref8}. Interestingly, the onset of DM1 leads before depressive episodes for male patients, while the onset of DM2 lags behind schizoaffective disorders and schizophrenia for females. This is consistent with findings in the literature, for example symptoms of depression and anxiety have been shown to be risk factors for the onset of DM2 independent of established risk factors for diabetes in a prospective cohort study, while for DM1 no such relation was identified \cite{ref9}. Therefore DM1 patients may become depressed due to complications and the burden of the disease, while for schizoaffective disorders in DM2 patients there may be a similar pathogenetic mechanism or medication-associated increased risk with gender-differences in the strength of these associations. Depressive and schizoaffective disorders tend to show higher relative risks for DM2 than DM1 patients.

{\bf Diseases of the nervous system.} Vascular dementia and Alzheimer are classified as being lead by diabetes \cite{ref23}. Both show an excess of female patients. The relationship between Parkinson's disease (PD) and diabetes is disputed. There were two large prospective studies finding an increased risk for PD associated with diabetes, one study finding no association, and one study reporting lower risk of diabetes \cite{ref24}. Here PD is comorbid with an excess of male patients, it was detected as being lead by DM2 for males. There is a known association between epilepsy and DM1, especially in youth \cite{ref25}. For ages below 25 we find an excess of female patients which switches to an excess of males in later years. The relative risk is also significant for DM2 (4.6, CI 3.1-6.9 for DM1 and 1.6, CI 1.4-1.7 for DM2). The increased risk in young type 1 diabetics may be linked to ketoacidosis as a 2 fold higher risk of epilepsy was found in children and adolescents with metabolic acidosis \cite{ref26}. A 4fold greater risk of DM1 was also described in young adults with epilepsy \cite{ref27}. Both metabolic extremes -- hypoglycemia and diabetic ketoacidosis -- relate to EEG abnormalities in diabetic children which may increase risk of epilepsy.

Other comorbidities related to the nervous system comprise sleep disorders including sleep apnoe (dominated by males), transient cerebral ischemic attacks, and facial nerve disorders. 

{\bf Retinopathies.} Retinopathies are well known comorbidities \cite{ref10}. Cataracts, retinal detachments, glaucoma, disorders of the vitreous body, and blindness are identified here. Relative risks tend to be higher for DM1 patients than for DM2, peaking at ages 45-70.

{\bf Circulatory diseases.} Primary hypertension is a comorbidity \cite{ref11} with relative risks of 5.3 (CI 4.8-5.9) for DM1 and 9.5 (CI 8.8-10) DM2. For ages 20-40 males dominate, before and after this window female patients outweigh male ones. Other comorbid diseases of the circulatory system show a consistent excess of male patients, including ischemic, pulmonary, and other heart diseases (cardiomyopathy, valvular disorders, tachycardia), as well as cerebrovascular diseases and diseases of the arteries and veins \cite{ref5, ref7, ref28}. For male patients DM2 leads chronic ischemic heart disease, for females DM2 leads heart failure, for both genders it leads atrial fibrillation and flutter. On the other hand, DM2 lags behind the onset of Angina pectoris (females) and nonrheumatic mitral valve disorders (males).
Benign pleural effusion, representing a symptom of various underlying diseases is dominated by males and classified as being lead by DM2 for females. In diabetic patients it may be related to left ventricular dysfunction as described previously \cite{ref29}. 

{\bf Respiratory diseases.} Pneumonia and acute bronchitis show increased relative risks for older ages (e.g. for pneumonia 2.7, CI 2.4-3.0, for DM1, 2.3, CI 2.1-2.4, for DM2). Diabetes is often identified as independent risk factor for lower respiratory tract infections \cite{ref30}. Other chronic obstructive pulmonary diseases (COPD) are dominated by males and classified as being lead by DM2 for females. Individuals with COPD are substantially more likely to have pre-existing DM \cite{ref31}, on the other hand lung function impairment in COPD is a risk factor for developing diabetes and insulin resistance \cite{ref32}.

{\bf Diseases of the digestive system.} Gastritis and non-infective gastroenteritis are comorbidities with a female excess, the latter being lead by DM2. For inguinal hernia there are decreased relative risks irrespective of patient age (0.37, CI 0.23-0.60, for DM1 and 0.48, CI 0.41-0.56, for DM2). Alcoholic and toxic liver diseases, as well as cystic fibrosis and pancreatitis are comorbidities \cite{ref33}, with a male excess. Cholelithiasis is a female dominated comorbidity, a known association which has been claimed to disappear once one controls for obesity \cite{ref34}, where there is an increased hepatic secretion of cholesterol. Intestinal malabsorption (including celiac disease) shows elevated risks for ages 10-25 for DM1 (10, CI 6.3-17) with a weak female excess, it also leads the onset of DM1 for females; there is no significant association with DM2. In particular, there is a known association between DM1 and celiac disease \cite{ref35}. 

{\bf Diseases of the skin.} Pressure and non-pressure ulcers exhibit higher risks for DM1 (7.2, CI 5.2-9.9, and 7.4, CI 5.8-9.4, respectively) than DM2 patients (2.2, CI 2.0-2.4 and 4.2, CI 3.9-4.6), which may be related to the diabetic foot ulcer \cite{ref5}. They are lead by DM2 for both males and females.

{\bf Diseases of the musculoskeletal system.} For spondylosis and dorsalgia elevated risks are dominated by females. Acquired deformities of fingers and toes, and internal derangement of knees show reduced risks for both DM1 and DM2 over the entire age range 15-80, which cannot be interpreted without further information on the patients' histories and physical exams.

{\bf Diseases of the genitourinary system.} Chronic kidney diseases and acute kidney diseases are identified as comorbidities \cite{ref5}, along with the nephrotic syndrome and glomerular disorders, each with an excess of male patients. For males there are increased risks for disorders of prepuce, while for females there is increased risk for disorders of the urinary system. Although urologic complications have been recognized since a long time, little is known about DM as a pathophysiological risk factor for development of lower urinary tract symptoms (LUTS) in women \cite{ref36}. Evidence from epidemiological studies suggests that asymptomatic bacteriuria (ASB) and symptomatic urinary tract infections occur more commonly in women with DM compared to non-diabetic controls \cite{ref37}. Prevalence data of urinary incontinence in diabetics are limited, recent evidence suggests an increased prevalence of this condition among women with DM2 \cite{ref38} and large observational studies have identified urge incontinence as increased among women with DM \cite{ref39}.

{\bf Limitations.} Only persons with inpatient stays were included in the study. To test if this pre-selection introduces a bias in our results, we repeated the study with a sample of all patients having been prescribed at least once a drug used in diabetes (ATC code starting with 'A10') in 2006 or 2007 at least once. We compare the co-occurrences of their diseases with the incidences in the rest of the population. This assumes that DM patients with no hospital stay in the study period have no diagnosis and therefore no comorbidities. Although this is a highly incorrect assumption, it serves as a conservative test-assumption, which allows to test if the comorbidities are simply significant as a consequence of our limited inpatient sample, where the number of diagnoses per patients is higher than what would be expected from the entire population. Results are shown in the supplement in Figure S1. There only one out of the 123 comorbidities using the inpatient sample has a $p$-value greater than 0.05 (M23), all other remain significant ($p<0.05$). Significance of comorbidity in the inpatient sample is therefore highly representative of the entire population. Other limitations relate to the coding quality of disorders in the medical claims data, this has been shown to lead to an under-reporting of comorbidities \cite{ref40} and may cause false negatives in our testing procedure.

In summary, in this work we developed a standardized testing procedure to obtain a complete comorbidity profile for DM1 and DM2 using medical claims data. Out of 39,938 possible comorbidities, we identified 123 with significantly increased or decreased risks as functions of patient age and gender. Comorbidities are investigated by a lead/lag analysis of the onset of the two disorders. We recover all the well known diabetic comorbidities (retinopathies, hypertension, overweight, chronic kidney diseases, etc) and find a number of lesser known comorbidities such as epilepsy, sepsis, or mental disorders. This shows the validity and power of this methodological approach, which may be readily applied to carry out systematic comorbidity analyses for other chronic disorders.

\clearpage

\setcounter{figure}{0}
\setcounter{table}{0}

\section{Supplementary Information for 'Quantifying age- and gender-related diabetes comorbidity risks using nation-wide big claims data'}

\subsection{Description of Indicators}

{\bf Gender ratio.} The number of male/female DM1 and DM2 patients in age group t is denoted $N_{m/f}(d,t)$, the number of male/female patients who also have diagnoses $x$ as diabetic comorbidity is $N_{m/f}(d,x,t)$.
The gender ratio $GR(x,t)$ is then related to the logarithmic quotient of the percentage of female and male diabetes patients who also have diagnoses $x$,
\begin{equation}
	GR(x,t) = \log \left( \frac{1+\frac{N_f(d,x,t)}{N_f(d,t)}}{1+\frac{N_m(d,x,t)}{N_m(d,t)}} \right) \quad.	
\label{equ1}	
\end{equation}	
Further, let $N_{m/f}(t)$ be the number of male/female patients in age group $t$, and $N_{m/f}(y,t)$ the number of male/female patients having $y$ diagnoses, or being described $y$ drugs, in age group $t$.
The gender ratios for drugs and diagnoses $GR(y,t)$ are obtained by replacing $N_{m/f}(d,t)$ ($N_{m/f}(d,x,t)$) by $N_{m/f}(t)$ ($N_{m/f}(y,t)$) in equation \ref{equ1}.

{\bf Lead/lag indicator.} $Let M(d_i,x,t)$ be the number of patients diagnosed with $d_i$ and $x$ in year $t$. To indicate the number of patients not having $d_i$ and/or $x$, we exchange $d_i$ ($x$) by $\neg d_i$ ($\neg x$) respectively, for example $M(d_i,\neg x,t)$ is the number of people having diabetes but not diagnoses $x$ in year $t$.
The patient numbers $M(d_i,x,t_2 |d_i,x,t_1)$ denote the number of patients having diagnoses $d_i$ and $x$ in year $t_2$ and $t_1$. The indicator for $d_i$ leading $x$, $I_{lead}(d_i,x)$, is then given by
\begin{eqnarray}
	I_{lead}(d_i,x) & =  & \frac{M(d_i,x,2007|d_i,\neg x,2006)}{M(d_i,x,2007)} \\ \nonumber
 & - & \frac{M(d_i,x,2006|d_i,\neg x,2007)}{M(d_i,x,2006)}	\quad.
\label{equ2}	
\end{eqnarray}	
The first summand is the number of patients which had both $d_i$ and $x$ in 2007, but only $d_i$ in 2006. This is compared to the number of patients having both $d_i$ and $x$ in 2006, but only $d_i$ in 2007. 
If there would be any temporal order between having $x$ and $d_i$, for example that one is much more likely to be diagnosed with $x$ if one already has the diagnoses $d_i$, these two terms can be expected to substantially differ. In particular, values for $I_{lead}(d_i,x)>0$ suggest that $d_i$ leads $x$.
Using the same rational, an indicator quantifying whether being diagnosed with $d_i$ tends to occur after one has been diagnosed with $x$ can be constructed, the lag indicator $I_{lag}(d_i,x)$ given by
\begin{eqnarray}
	I_{lag}(d_i,x) & =  & \frac{M(d_i,x,2007|\neg d_i,x,2006)}{M(d_i,x,2007)} \\ \nonumber
 & - & \frac{M(d_i,x,2006|\neg d_i, x,2007)}{M(d_i,x,2006)}	\quad.
\label{equ3}	
\end{eqnarray}	
Positive values of $I_{lag}(d_i,x)$ suggest that there is a tendency that the diagnoses of diabetes lags the diagnoses of the other disease $x$.
One may observe high values for $I_{lead/lag}(d_i,x)$ because the frequency of the diagnoses $x$ itself is very small and thus the quotients in the above definitions become large. To compensate for this we exclude diagnoses $x$ for which $M(d_i,x,2007)<z$ for some value $z$. 
Finally a statistical test is developed to assess the significance of $I_{lead/lag}(d_i,x)$ values. 
First, surrogate data is created by keeping the diagnoses for each patient fixed, but scrambling the information about the year when the diagnoses was made. 
Assume that patient a has $N_a$ diagnoses $\{ x_i \}_{i \in \{1,\dots,N_a \} }$ made in the years $\{ \tau_i \}_{i \in \{1,\dots,N_a \} }$ respectively. 
The surrogate data is constructed by replacing $\{ \tau_i \}_{i \in \{1,\dots,N_a \} }$ by a random permutation of itself and then compute the surrogate indicators $\tilde I_{lead/lag}(d_i,x)$. 
This procedure is repeated 100 times. 
Let $\tilde M(d_i,x,2007)$ be the number of patients with diagnoses $d_i$ and $x$ in the surrogate data. For each value of $I_{lead/lag}(d_i,x)$ we pick all surrogate indicator values for which 
\begin{equation}
\left| \frac{\tilde M(d_i,x,2007) - M(d_i,x,2007)}{\tilde M(d_i,x,2007)} \right| <0.5	 
\label{equ4}	
\end{equation}	
holds to avoid spurious results from potential size-dependences on the frequencies of the diagnoses. 
The $p$-value for $I_{lead/lag}(d_i,x)$  is the fraction of surrogate values for which both $I_{lead/lag}(d_i,x) > \tilde I_{lead/lag}(d_i,x)$ and equation \ref{equ4} holds. A $p$-value of 0.01, for instance, indicates that one percent of the relevant surrogate indicator values are larger than the observed lead or lag indicator; there is a chance of one percent to obtain indicator values as large as the observed ones from a patient data-set, where all time information has been randomly shuffled. 
The above method is applied to male and female patients with Type I (using $z=3$) and II ($z=20$) diabetes, respectively. For DM1 only patients below an age of thirty were included in the analysis.

\begin{figure*}[tbp]
 \begin{center}
 \includegraphics[width=165mm]{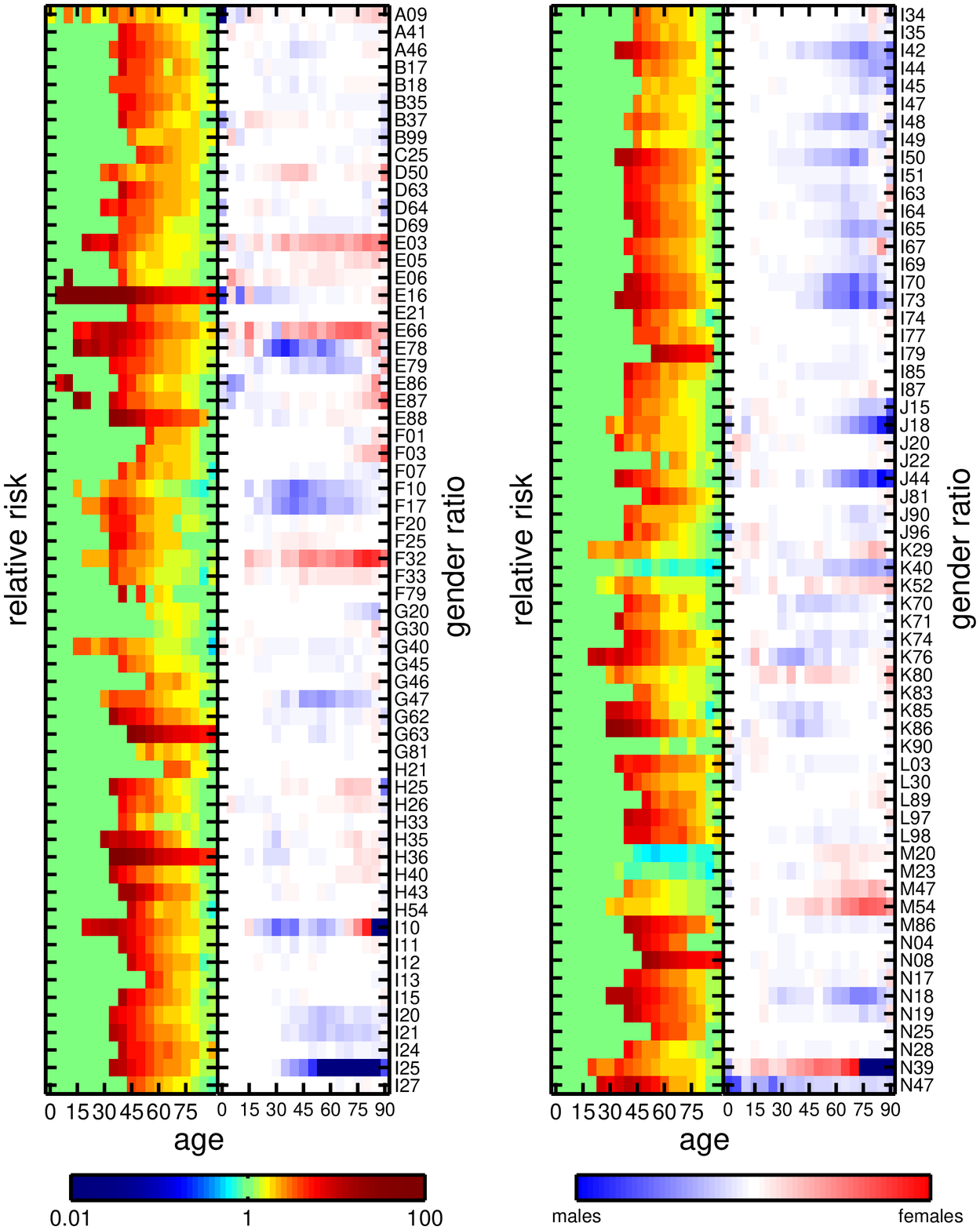}
  \end{center}
 \caption{Values for relative risks (left panels) and gender ratios (right panels) for the 123 identified comorbidities computed using a patient sample of all persons receiving a drug used in diabetes (ATC code A10). The results from the inpatient sample are reproduced to large parts, only disorder M23 exhibits non-significant $p$-values.}
  \label{FigS1}
 \end{figure*}

%\begin{table*}[p]
%\clearpage

\begin{center}
\begin{longtable*}{ll|lll|lll}
\caption[ICD code and disease name for the 123 comorbidities identified in the co-occurrence analysis. For the age groups with the smallest $p$-value the relative risks $RR$, patient ages, and the corresponding $p$-values are shown for DM1 and DM2, respectively. Where the patient sample was too small to apply the statistical tests missing values are shown.]{ICD code and disease name for the 123 comorbidities identified in the co-occurrence analysis. For the age groups with the smallest $p$-value the relative risks $RR$, patient ages, and the corresponding $p$-values are shown for DM1 and DM2, respectively. Where the patient sample was too small to apply the statistical tests missing values are shown.}\\
\hline
 & & & DM1 & & & DM2 & \\
ICD & Diagnosis & $p$ & $RR_1$& age & $p$ & $RR_2$& age \\
\hline
\endfirsthead
\multicolumn{8}{c}%
{{\bfseries \tablename\ \thetable{} -- continued from previous page}} \\
\hline
 & & & DM1 & & & DM2 & \\
ICD & Diagnosis & $p$ & $RR_1$& age & $p$ & $RR_2$& age \\
\hline
\endhead
\hline \multicolumn{8}{r}{\textit{Continued on next page}} \\
\endfoot
\hline
\endlastfoot
A09 & Diarrhoea and gastroenteritis of infectious origin &	0.004 &	4.4 (3.0-6.3) &	45-50 &	0.0003 &	1.6 (1.5-1.7) &	80-85 \\
A41	& Other septicaemia &	$<10^{-4}$ &	12 (8.2-18) &	45-50 &	$<10^{-4}$ &	2.7 (2.4-2.9) &	65-70 \\
A46 &	Erysipelas &	0.0009 &	3.7 (2.9-4.8) &	65-70 &	$<10^{-5}$ &	2.7 (2.5-3.0) &	65-70 \\
B17 &	Other acute viral hepatitis &	0.006 &	7.1 (4.0-13) &	50-55 &	0.002 &	6.9 (4.5-11) &	40-45 \\
B18 &	Chronic viral hepatitis &	0.0007 &	6.3 (4.4-9.0) &	50-55 &	0.0006 &	3.2 (2.6-3.8) &	50-55 \\
B35 &	Dermatophytosis &	0.01 &	3.5 (2.4-5.2) &	65-70 &	0.0005 &	2.6 (2.2-3.0) &	65-70 \\
B37 &	Candidiasis &	0.002 &	5.5 (3.7-8.2) &	55-60 &	0.002 &	2.1 (1.8-2.5) &	65-70 \\
B99 &	Other and unspecified infectious diseases &	0.003 &	4.7 (3.3-6.9) &	65-70 &	0.004 &	1.7 (1.5-2.0) &	80-85 \\
C25 &	Malignant neoplasm of pancreas &	0.0006 &	8.6 (5.6-13) &	55-60 &	0.0006 &	2.5 (2.1-2.8) &	65-70 \\
D50 &	Iron deficiency anaemia &	0.0003 &	3.7 (3.0-4.6) &	70-75 &	$<10^{-5}$ &	2.7 (2.4-2.8) &	65-70 \\
D63 &	Anaemia in chronic disease &	$<10^{-4}$ &	6.3 (4.9-8.2) &	65-70 &	$<10^{-4}$ &	2.8 (2.5-3.2) &	65-70 \\
D64 &	Other anaemia &	$<10^{-5}$ &	4.9 (4.2-5.8) &	65-70 &	$<10^{-6}$ &	2.6 (2.4-2.8) &	65-70 \\
D69 & Purpura and other haemorrhagic conditions &	0.009 &	3.6 (2.4-5.4) &	55-60 &	0.001 &	2.0 (1.7-2.2) &	65-70 \\
E03 &	Other hypothyroidism &	0.0001 &	9.3 (6.4-13) &	35-40 &	0.0001 &	1.7 (1.6-1.9) &	65-70 \\
E05 &	Thyrotoxicosis &	0.009 &	5.3 (3.1-8.9) &	40-45 &	0.001 &	1.6 (1.5-1.8) &	70-75 \\
E06 &	Thyroiditis &	$<10^{-6}$ &	38 (26-57) &	10-15 &	0.04 &	1.7 (1.3-2.1) &	55-60 \\
E16 &	Other disorders of pancreatic internal secretion &	$<10^{-15}$ &	170 (130-223) &	20-25 &	$<10^{-8}$ &	5.4 (5.0-5.7) &	80-85 \\
E21 & Hypoparathyroidism &	0.0006 &	6.1 (4.4-8.6) &	65-70 &	0.0002 &	3.0 (2.6-3.5) &	67-70 \\
E66 &	Obesity &	$<10^{-6}$ &	3.7 (3.3-4.1) &	65-70 &	$<10^{-16}$ &	4.8 (4.6-5.0) &	55-60 \\
E78 &	Dis. of lipoprotein metabolism and other lipidaemia &	$<10^{-5}$ &	3.4 (3.0-3.7) &	60-65 &	$<10^{-15}$ &	3.9 (3.7-4.0) &	55-60 \\
E79 &	Dis. or purine and pyrimidine metabolism &	$<10^{-4}$ &	3.2 (2.8-3.7) &	65-70 &	$<10^{-7}$ &	2.5 (2.4-2.6) &	65-70 \\
E86 &	Volume depletion &	0.0001 &	9.4 (6.5-14) &	$<10^{-15}$ &	0.0003 &	1.7 (1.6-1.8) &	75-80 \\
E87 &	Other dis. of fluid, electrolyte and acid-base balance &	$<10^{-4}$ &	5.9 (4.7-7.3) &	60-65 &	$<10^{-5}$ &	2.5 (2.3-2.6) &	65-70 \\
E88 &	Other metabolic disorders &	0.0001 &	6.4 (4.9-8.4) &	60-65 &	$<10^{-7}$ &	5.1 (4.7-5.6) &	60-65 \\
F01 &	Vascular dementia &	0.0008 &	3.0 (2.4-3.6) &	75-80 &	$<10^{-4}$ &	2.1 (1.9-2.2) &	75-80 \\
F03 &	Unspecified dementia &	0.001 &	2.5 (2.1-2.9) &	75-80 &	$<10^{-4}$ &	1.8 (1.7-1.9) &	75-80 \\
F07 &	Pers. and behav. dis. due to known physiological cond. &	0.002 &	3.8 (2.9-5.1) &	75-80 &	0.001 &	2.4 (2.0-2.7) &	65-70 \\
F10 &	Use of alcohol &	0.02 &	2.3 (1.7-3.2) &	40-45 &	0.0007 &	2.1 (1.9-2.4) &	45-50 \\
F17 &	Use of tobacco &	0.0004 &	3.3 (2.7-4.1) &	50-55 &	$<10^{-6}$ &	2.8 (2.6-3.0) &	50-55 \\
F20 &	Schizophrenia &	0.09 &	2.1 (1.2-3.7) &	65-70 &	0.001 &	3.4 (2.7-4.2) &	45-50 \\
F25 &	Schizoaffective disorders &	- &	- &	- &	0.003 &	4.3 (3.0-6.2) &	40-45 \\
F32 &	Depressive episode &	0.001 &	2.3 (1.9-2.6) &	65-70 &	$<10^{-4}$ &	1.7 (1.6-1.8) &	65-70 \\
F33 &	Recurrent depressive disorder &	0.05 &	1.9 (1.3-2.6) &	65-70 &	0.003 &	4.8 (3.3-7.0) &	35-40 \\
F79 &	Unspecified mental retardation &	- &	- &	- &	0.009 &	9.3 (6.0-15) &	40-45 \\
G20 &	Parkinson's disease &	0.002 &	2.3 (1.9-2.7) &	75-80 &	0.001 &	1.5 (1.4-1.6) &	75-80 \\
G30 &	Alzheimer's disease &	0.04 &	1.7 (1.3-2.1) &	75-80 &	0.001 &	1.5 (1.4-1.7) &	75-80 \\
G40 &	Epilepsy &	0.004 &	4.6 (3.1-6.9) &	35-40 &	0.003 &	1.6 (1.4-1.7) &	65-70 \\
G45 &	Transient cerebral ischaemic attacks &	0.006 &	2.8 (2.1-3.7) &	60-65 &	0.0005 &	1.7 (1.6-1.9) &	65-70 \\
G46 &	Vascular syndromes of brain in cerebrovascular diseases &	0.002 &	3.9 (3.0-5.3) &	80-85 &	0.012 &	1.9 (1.6-2.3) &	70-75 \\
G47 &	Sleep disorders &	0.02 &	1.9 (1.5-2.4) &	60-65 &	0.0001 &	2.3 (2.1-2.6) &	50-55 \\
G62 &	Other polyneuropathies &	$<10^{-5}$ &	8.2 (6.6-10) &	55-60 &	$<10^{-5}$ &	2.7 (2.5-3.0) &	65-70 \\
G63 &	Polyneuropathy in diseases c. e. &	$<10^{-10}$ &	20 (17-23) &	65-70 &	$<10^{-11}$ &	8.0 (7.6-8.4) &	65-70 \\
G81 &	Hemiplegia &	0.008 &	3.9 (2.6-5.3) &	65-70 &	0.001 &	1.9 (1.6-2.3) &	65-70 \\
H21 &	Other disorders of iris and ciliary body &	0.001 &	10 (6.2-16) &	75-80 &	0.006 &	6.2 (4.0-9.4) &	55-60 \\
H25 &	Senile cataract &	$<10^{-4}$ &	4.7 (3.9-5.7) &	55-60 &	0.0002 &	2.0 (1.8-2.2) &	55-60 \\
H26 &	Other cataract &	0.001 &	5.2 (3.7-7.3) &	55-60 &	0.004 &	2.3 (1.9-2.8) &	55-60 \\
H33 &	Retinal detachments and breaks &	0.0003 &	18 (10-32) &	30-35 &	0.05 &	2.3 (1.6-3.5) &	45-50 \\
H35 &	Other retinal disorders &	$<10^{-6}$ &	10 (8-12) &	55-60 &	$<10^{-6}$ &	4.3 (3.9-4.7) &	55-60 \\
H36 &	Retinal disorders in diseases c. e. &	$<10^{-16}$ &	200 (160-250) &	30-35 &	0.08 &	230 (180-310) &	30-35 \\
H40 &	Glaucoma &	0.0002 &	14 (8.7-23) &	40-45 &	0.003 &	1.9 (1.6-2.2) &	60-65 \\
H43 &	Disorders of vitreous body &	$<10^{-6}$ &	36 (25-53) &	45-50 &	$<10^{-4}$ &	4.9 (4.1-5.9) &	55-60 \\
H54 &	Blindness and low vision &	0.01 &	4.6 (2.7-7.8) &	65-70 &	0.002 &	3.0 (2.4-3.6) &	65-70 \\
I10 &	Essential (primary) hypertension &	$<10^{-8}$ &	5.3 (4.8-5.9) &	65-70 &	$<10^{-16}$ &	9.5 (8.8-10) &	45-50 \\
I11 &	Hypertensive heart disease &	0.0005 &	3.6 (2.9-4.5) &	60-65 &	$<10^{-7}$ &	2.9 (2.7-3.1) &	65-70 \\
I12 &	Hypertensive chronic kidney disease &	0.0002 &	20 (12-35) &	45-50 &	0.001 &	3.4 (2.8-4.1) &	65-70 \\
I13 &	Hypertensive heart dis. and hypertensive renal dis. &	0.002 &	7.6 (4.7-12) &	65-70 &	0.001 &	5.1 (3.9-6.6) &	60-65 \\
I15 &	Secondary hypertension &	0.01 &	4.8 (2.7-8.3) &	55-60 &	0.0006 &	11 (7.1-18) &	40-45 \\
I20 &	Angina pectoris &	0.0007 &	3.2 (2.6-3.9) &	55-60 &	$<10^{-7}$ &	2.3 (2.2-2.5) &	65-70 \\
I21 &	Acute myocardial infarction &	$<10^{-4}$ &	3.2 (2.8-3.7) &	75-80 &	$<10^{-7}$ &	2.9 (2.7-3.1) &	65-70 \\
I24 &	Other acute ischaemic heart diseases &	$<10^{-4}$ &	6.6 (5.2-8.3) &	65-70 &	$<10^{-4}$ &	3.1 (2.8-3.4) &	65-70 \\
I25 &	Chronic ischaemic heart disease &	$<10^{-8}$ &	5.3 (4.7-5.9) &	60-65 &	$<10^{-16}$ &	3.3 (3.2-3.4) &	65-70 \\
I27 &	Other pulmonary heart diseases &	0.004 &	2.5 (2.0-3.1) &	75-80 &	0.0002 &	2.4 (2.1-2.6) &	65-70 \\
I34 &	Nonrheumatic mitral valve disorders &	0.001 &	2.4 (2.0-2.8) &	75-80 &	$<10^{-5}$ &	2.3 (2.2-2.5) &	65-70 \\
I35 &	Nonrheumatic aortic valve disorders &	0.004 &	2.7 (2.1-3.4) &	65-70 &	$<10^{-4}$ &	1.8 (1.7-1.9) &	75-80 \\
I42 &	Cardiomyopathy &	$<10^{-4}$ &	4.0 (3.3-4.7) &	65-70 &	$<10^{-8}$ &	3.0 (2.8-3.2) &	65-70 \\
I44 &	Atrioventricular and left bundle-branch block &	0.0004 &	5.1 (3.9-6.8) &	60-65 &	$<10^{-5}$ &	2.5 (2.3-2.7) &	65-70 \\
I45 &	Other conduction disorders &	0.007 &	3.1 (2.3-4.2) &	70-75 &	0.0007 &	2.1 (1.9-2.4) &	70-75 \\
I47 &	Paroxysmal tachycardia &	0.0004 &	3.3 (2.7-4.0) &	75-80 &	0.0003 &	2.0 (1.8-2.2) &	65-70 \\
I48 &	Atrial fibrillation and flutter &	$<10^{-4}$ &	2.1 (1.9-2.3) &	75-80 &	$<10^{-9}$ &	2.0 (1.9-2.0) &	75-80 \\
I49 &	Other cardiac arrhythmias &	0.008 &	2.2 (1.8-2.8) &	70-75 &	0.001 &	1.5 (1.4-1.6) &	75-80 \\
I50 &	Heart failure &	$<10^{-7}$ &	5.2 (4.7-5.9) &	65-70 &	$<10^{-14}$ &	3.8 (3.6-3.9) &	65-70 \\
I51 &	Complications \& ill-defined descriptions of heart dis. &	0.0002 &	3.4 (2.8-4.0) &	75-80 &	$<10^{-6}$ &	3.1 (2.8-3.4) &	65-70 \\
I63 &	Cerebral Infarction &	0.0002 &	4.4 (3.5-5.6) &	60-65 &	$<10^{-6}$ &	2.6 (2.4-2.8) &	65-70 \\
I64 &	Stroke, not specified as haemorrhage or infarction &	$<10^{-4}$ &	4.5 (3.7-5.5) &	65-70 &	$<10^{-6}$ &	2.8 (2.6-3.1) &	65-70 \\
I65 &	Occlusion and stenosis of precerebral arteries &	$<10^{-4}$ &	5.3 (4.4-6.4) &	60-65 &	$<10^{-8}$ &	3.0 (2.8-3.1) &	65-70 \\
I67 &	Other cerebrovascular diseases &	0.0001 &	5.0 (3.9-6.2) &	60-65 &	$<10^{-5}$ &	2.2 (2.1-2.3) &	70-75 \\
I69 &	Sequelae of cerebrovascular disease &	0.005 &	3.5 (2.5-4.9) &	60-65 &	$<10^{-6}$ &	3.0 (2.8-3.3) &	65-70 \\
I70 &	Atherosclerosis &	$<10^{-8}$ &	6.0 (5.4-6.7) &	70-75 &	$<10^{-11}$ &	3.0 (2.5-3.5) &	65-70 \\
I73 &	Other peripheral vascular diseases &	$<10^{-7}$ &	5.7 (5.1-6.5) &	65-70 &	$<10^{-13}$ &	3.9 (3.7-4.1) &	65-70 \\
I74 &	Arterial embolism and thrombosis &	0.002 &	5.4 (3.7-7.7) &	60-65 &	0.0004 &	3.0 (2.5-3.5) &	60-65 \\
I77 &	Other disorders of arteries and arterioles &	0.001 &	10 (6.1-17) &	50-55 &	0.0006 &	2.5 (2.2-2.9) &	70-75 \\
I79 &	D. of arteries, arterioles and capillaries in diseases c. e. &	$<10^{-7}$ &	23 (18-29) &	70-75 &	$<10^{-6}$ &	5.7 (5.2-6.1) &	75-80 \\
I85 &	Esophageal varices &	0.005 &	4.3 (2.9-6.4) &	55-60 &	0.0003 &	3.0 (2.5-3.5) &	55-60 \\
I87 &	Other disorders of veins &	0.009 &	3.7 (2.5-5.5) &	60-65 &	0.0004 &	2.3 (2.1-2.6) &	65-70 \\
J15 &	Bacterial pneumonia n. e. c. &	0.02 &	2.5 (1.8-3.3) &	70-75 &	0.0001 &	2.4 (2.2-2.7) &	65-70 \\
J18 &	Pneumonia, organism unspecified &	$<10^{-4}$ &	2.7 (2.4-3.0) &	75-80 &	$<10^{-6}$ &	2.3 (2.1-2.4) &	65-70 \\
J20 &	Acute bronchitis &	0.03 &	1.9 (1.5-2.6) &	80-85 &	0.0007 &	2.2 (2.0-2.5) &	65-70 \\
J22 &	Unspecified acute lower respiratory infection &	0.02 &	3.4 (2.1-5.6) &	75-80 &	0.002 &	3.1 (2.5-3.9) &	65-70 \\
J44 &	Other chronic pulmonary diseases &	0.0003 &	2.9 (2.5-3.5) &	55-60 &	$<10^{-8}$ &	2.2 (2.1-2.3) &	65-70 \\
J81 &	Pulmonary oedema &	0.0003 &	9.9 (6.6-15) &	60-65 &	$<10^{-4}$ &	3.7 (3.2-4.3) &	65-70 \\
J90 &	Pleural effusion, not elsewhere classified &	0.0004 &	3.4 (2.1-5.6) &	75-80 &	0.0001 &	3.1 (2.5-3.9) &	65-70 \\
J96 &	Respiratory failure, n. e. c. &	0.001 &	9.4 (5.7-16) &	10-15 &	$<10^{-4}$ &	2.3 (2.1-2.5) &	65-70 \\
K29 &	Gastritis and duodenitis &	0.01 &	1.8 (1.5-2.1) &	60-65 &	0.0002 &	1.5 (1.4-1.5) &	70-75 \\
K40 &	Inguinal hernia &	0.04 &	0.37 (0.23-0.60) &	75-80 &	0.001 &	0.48 (0.41-0.56) &	65-70 \\
K52 &	Other \& unspec. noninfective gastroenteritis and colitis &	0.006 &	2.1 (1.7-2.6) &	75-80 &	0.0008 &1.5 (1.4-1.6) &	80-85 \\
K70 &	Alcoholic liver disease &	0.005 &	4.0 (2.7-5.7) &	45-50 &	0.0001 &	2.6 (2.3-2.9) &	55-60 \\
K71 &	Toxic liver disease &	0.05 &	2.8 (1.6-4.9) &	55-60 &	0.0003 &	14 (8.5-23) &	35-40 \\
K74 &	Fibrosis and cirrhosis of liver &	0.0004 &	5.0 (3.7-6.6) &	55-60 &	$<10^{-4}$ &	2.4 (2.2-2.7) &	65-70 \\
K76 &	Other diseases of liver &	0.0005 &	2.7 (2.2-3.1) &	60-65 &	$<10^{-8}$ &	3.0 (2.8-3.2) &	55-60 \\
K80 &	Cholelithiasis &	0.008 &	1.7 (1.5-2.0) &	75-80 &	$<10^{-4}$ &	1.5 (1.4-1.6) &	80-85 \\
K83 &	Other diseases of biliary tract &	0.003 &	7.5 (4.6-12) &	50-55 &	0.03 &	1.4 (1.2-1.6) &	80-85 \\
K85 &	Acute pancreatitis &	0.006 &	4.5 (3.0-6.9) &	50-55 &	$<10^{-4}$ &	4.7 (3.9-5.8) &	45-50 \\
K86 &	Other diseases of pancreas &	$<10^{-5}$ &	13 (9.3-17) &	50-55 &	$<10^{-5}$ &	11 (8.4-14) &	40-45 \\
K90 &	Intestinal malabsorption &	0.001 &	10 (6.3-17) &	10-15 &	0.4 &	1.2 (0.7-2.2) &	60-65 \\
L03 &	Cellulitis and acute lymphangitis &	0.0005 &	5.5 (4.0-7.4) &	65-70 &	0.0006 &	2.5 (2.1-2.9) &	65-70 \\
L30 &	Other and unspecified dermatitis &	0.01 &	5.1 (3.0-8.7) &	45-50 &	0.001 &	1.9 (1.7-2.1) &	80-85 \\
L89 &	Decubitus ulcer &	0.0002 &	7.2 (5.2-9.9) &	65-70 &	$<10^{-4}$ &	2.2 (2.0-2.4) &	80-85 \\
L97 &	Ulcer of lower limb n. e. c. &	$<10^{-4}$ &	7.4 (5.8-9.4) &	65-70 &	$<10^{-6}$ &	4.2 (3.9-4.6) &	65-70 \\
L98 &	Other disorders of skin and subcutaneous tissue &	$<10^{-5}$ &	9.0 (7.2-11) &	70-75 &	$<10^{-4}$ &	3.4 (3.1-3.9) &	65-70 \\
M20 &	Acquired deformities of fingers and toes &	0.09 &	0.44 (0.25-0.88) &	60-65 &	0.002 &	0.41 (0.34-0.49) &	65-70 \\
M23 &	Internal derangement of knee &	0.03 &	0.31 (0.19-0.52) &	60-65 &	0.0006 &	0.45 (0.40-0.52) &	65-70 \\
M47 &	Spondylosis &	0.04 &	1.7 (1.3-2.1) &	85-90 &	0.0003 &	1.6 (1.5-1.7) &	70-75 \\
M54 &	Dorsalgia &	0.06 &	1.3 (1.1-1.6) &	65-70 &	0.002 &	1.4 (1.3-1.5) &	60-65 \\
M86 &	Osteomyelitis &	$<10^{-6}$ &	13 (10-16) &	65-70 &	$<10^{-5}$ &	4.4 (3.9-5.0) &	65-70 \\
N04 &	Nephrotic syndrome &	$<10^{-4}$ &	33 (19-56) &	45-50 &	0.001 &	4.9 (3.8-6.4) &	60-65 \\
N08 &	Glomerular d. in diseases c. e. &	$<10^{-12}$ &	128 (98-166) &	40-45 &	$<10^{-9}$ &	8.6 (8.2-9.1) &	65-70 \\
N17 &	Acute renal failure &	$<10^{-4}$ &	13 (8.9-20) &	45-50 &	$<10^{-6}$ &	3.4 (3.1-3.7) &	65-70 \\
N18 &	Chronic renal failure &	$<10^{-10}$ &	8.0 (7.2-8.9) &	65-70 &	$<10^{-14}$ &4.2 (4.0-4.4) &	65-70 \\
N19 &	Unspecified renal failure &	$<10^{-7}$ &	9.4 (8.1-11) &	65-70 &	$<10^{-6}$ &	3.2 (2.9-3.5) &	65-70 \\
N25 &	Disorders resulting from impaired renal tubular fct. &	0.002 &	9.0 (5.5-15) &	70-75 &	0.001 &	3.9 (3.1-4.9) &	65-70 \\
N28	 &O. d. of kidney and ureter, n. e. c. &	0.004 &	5.8 (3.7-9.1) &	50-55 &	0.0008 &	2.7 (2.3-3.2) &	55-60 \\
N39 &	Other disorders of urinary system &	$<10^{-4}$ &	2.5 (2.2-2.8) &	80-85 &	$<10^{-7}$ &	1.8 (1.7-1.9) &	80-85 \\
N47 &	Disorders of prepuce &	0.007 &	6.0 (3.5-10) &	45-50 &	0.001 &	3.1 (2.5-3.8) &	55-60 \\
\end{longtable*}
\end{center}
%\label{crtab}
%\end{table*}

\begin{thebibliography}{99}

\bibitem{ref1}	Thurner S, Klimek P, Szell M, et al. Quantification of excess-risk for diabetes when born in times of hunger, in an entire population of a nation, across a century. {\it Proceedings of the National Academy of Sciences USA} 2013; {\bf 110}(12): 4703-7.

\bibitem{ref2}	Elixhauser A, Steiner C, Harris RD, et al. Comorbidity measures for use with administrative data. {\it Medical Care} 1998; {\bf 36}(1): 8-27.

\bibitem{ref3}	Van den Bussche H, Koller D, Kolonko T, et al. Which chronic diseases and disease combinations are specific to multimorbidity in the elderly? {\it BMC Public Health} 2011; {\bf 11}: 101-9.

\bibitem{ref4}	The Lancet. The diabetes pandemic. {\it The Lancet} 2011; {\bf 378}(9786): 99.

\bibitem{ref5}	Eko\'e JM, Rewers M, Williams R, Zimmet P. The Epidemiology of Diabetes Mellitus (Wiley, New Jersey, USA, 2008).

\bibitem{ref6}	Haffner SM, Lehto S, R\"onnemaa T, et al. Mortality from coronary heart disease in subjects with type 2 diabetes and in nondiabetec subjects with and without prior myocardial infarction. {\it New England Journal of Medicine} 1998; {\bf 339}: 229-34.

\bibitem{ref7}	Almdal T, Scharling H, Skov Jensen J, et al. The independent effect of type 2 diabetes mellitus on ischemic heart diesease, stroke and death. {\it Arch Intern Med} 2004; {\bf 164}(13): 1422-6.

\bibitem{ref8}	Anderson RJ, Freedland KE, Clouse RE, et al. The prevalence of comorbid depression in adults with diabetes. {\it Diabetes Care} 2001; {\bf 24}(6): 1069-78.

\bibitem{ref9}	Engum A. The role of depression and anxiety in onset of diabetes in a large population-based study. {\it Journal of Psychosomatic Research} 2007; {\bf 62}(1): 31-38.

\bibitem{ref10}	Fong DS, Aiello L, Gardner T, et al. Retinopathy in diabetes. {\it Diabetes Care} 2004; {\bf 27}: 584-7.

\bibitem{ref11}	Lago RM, Singh PP, Nesto RW. Diabetes and hypertension. {\it Nature Clinical Practice Endocrinology \& Metabolism} 2007; {\bf 3}: 667.

\bibitem{ref12}	Endel G. Health systems research in Austria: Part 1. {\it Social Security Online} 2011 (in German).

\bibitem{ref13}	Benjamini Y, Hochberg Y. Controlling the false discovery rate: a practical and powerful approach to multiple testing. {\it Journal of the Royal Statistical Society, Series B} 1995; {\bf 57}(1): 289-300.

\bibitem{ref14}	Mason AL, Lau JY, Hoang N, at al. Association of diabetes mellitus and chronic hepatitis C virus infection. {\it Hepatology} 1999; {\bf 29}(2): 328-33.

\bibitem{ref15}	Tan JS, Joseph WS. Common fungal infections of the feet in patients with diabetes mellitus. {\it Drugs \& Aging} 2004; {\bf 21}(2): 101-12.

\bibitem{ref16}	Schuetz P, Castro P, Shapiro N. Diabetes and Sepsis: Preclinical Findings and Clinical Relevance. {\it Diabetes Care} 2011; {\bf 34}(3): 771-8.

\bibitem{ref17}	Iskander KN, Osuchowski MF, Stearns-Kurosawa DJ, et al. Sepsis: Multiple abnormalities, heterogeneous responses, and evolving understanding. {\it Physiol Rev} 2013; {\bf 93}: 1247-1288.

\bibitem{ref18}	Everhart J, Wright D. Diabetes mellitus as a risk factor for pancreatic cancer. A meta-analysis. {\it JAMA} 1995; {\bf 273}: 1605-9.

\bibitem{ref19}	Pezzilli R, Pagano N, Is diabetes mellitus a risk factor for pancreatic cancer? {\it World J Gastroenterol} 2013; {\bf 19} (30): 4861-6.

\bibitem{ref20}	Al-Khoury S, Afzali B, Shah N, et al. Diabetes, kidney disease and anaemia: time to tackle a troublesome triad? {\it International Journal of Clinical Practice} 2007; {\bf 61}(2): 281-9.

\bibitem{ref21}	Vondra K, Vrbikova J, Dvorakova K. Thyroid gland diseases in adult patients with diabetes mellitus. {\it Minverva Endocrinol} 2005; {\bf 30}(4): 217-36.

\bibitem{ref22}	Baliunas DO, Taylor BJ, Irving H, et al. Alcohol as a risk factor for type 2 diabetes. {\it Diabetes Care} 2009; {\bf 32}(11): 2123-32.

\bibitem{ref23}	Ott A, Stolk RP, van Harskamp HA, et al. Diabetes mellitus and the risk of dementia. {\it Neurology 1999}; {\bf 53}(9): 1937-42.

\bibitem{ref24}	Wirdefeldt K, Adami HO, Cole P, et al. Epidemiology and etiology of Parkinson's disease: a review of the evidence. {\it Eur J Epidemiol} 2011; {\bf 26}: 1-58.

\bibitem{ref25}	O'Connel MA, Harvey S, Mackay MT, et al. Does epilepsy occur more frequently in children with type 1 diabetes? {\it Journal of Paediatrics and Child Health} 2008; {\bf 44}(10): 586-9.

\bibitem{ref26}	Schober E, Otto KP, Dost A, et al. Association of epilesy and type 1 diabetes mellitus in children and adolescents: is there an increased risk for diabetic ketoacidosis? {\it J Pediatr} 2012; {\bf 160}(4): 662-666.e1.

\bibitem{ref27}	McCorry D, Nicolson A, Smith D, et al. An association between type 1 diabetes and idiopathic generalized epilepsy. {\it Annals of Neurology} 2006; {\bf 59}: 204–206.

\bibitem{ref28}	Movahed M-R. Diabetes as a risk factor for cardiac conduction defects: a review. {\it Diabetes, Obesity and Metabolism} 2007; {\bf 9}(3):276-81.

\bibitem{ref29}	Chertow BS, Kadzielawa R, Burger AJ. Benign pleural effusions in long-standing diabetes mellitus. {\it Chest} 1991; {\bf 99}(5):1108-11.

\bibitem{ref30}	Koziel H, Koziel MJ. Pulmonary complications of diabetes mellitus. Pneumonia. {\it Infect Dis Clin North Am} 1995; {\bf 9}(1): 65-96.

\bibitem{ref31}	Feary JR, Rodrigues LC, Smith CJ, et al. Prevalence of major comorbidities in subjects with COPD and incidence of myocardial infarction and stroke: a comprehensive analysis using data from primary care. {\it Thorax} 2010; {\bf 65}(11): 956-62.

\bibitem{ref32}	Engstr\"om G, Hedblad B, Nilsson P, et al. Lung function, insulin resistance and incidence of cardiovascular disease: a longitudinal cohort study. {\it J Intern Med} 2003; {\bf 253}(5): 574-81.

\bibitem{ref33}	Garcia-Compean D, Jaquez-Quintana JO, Gonzalez-Gonzalez JA, et al. Liver cirrhosis and diabetes: Risk factors, pathophysiology, clinical implications and management. {\it World J Gatroenterol} 2009; {\bf 15}(3): 280-8.

\bibitem{ref34}	Bodmer M, Brauchli Y, Jick S, et al. Diabetes mellitus and the risk of cholecystectomy. {\it Digestive and Liver Disease} 2011; {\bf 43}(9), 742-7.

\bibitem{ref35}	Cronin CC, Shanahan F. Insulin-dependent diabetes mellitus and coeliac disease. {\it The Lancet} 1997; {\bf 349}(9058): 1096-7.

\bibitem{ref36}	Hill SR, Fayyad AM, Jones GR. Diabetes mellitus and female lower urinary tract symptoms: a review. {\it Neurourol Urodyn} 2008; {\bf 27}(5): 362-7.

\bibitem{ref37}	Stapleton A. Urinary tract infections in patients with DM. {\it Am J Med} 2002; {\bf 113}:80–4.

\bibitem{ref38}	Lifford KL, Curhan GC, Hu FB, et al. Type 2 diabetes and risk of developing urinary incontinence. {\it J Am Geriatr Soc} 2005; {\bf 53}: 1851–7.

\bibitem{ref39}	Jackson RA, Vittinghoff E, Kanaya AM, et al. Urinary incontinence in elderly women: Findings from the Health, Aging, and Body Composition Study. {\it Obstet Gynecol} 2004; {\bf 104}: 301–7.

\bibitem{ref40}	Quan H, Parsons G, Ghali WA. Validity of information on comorbidity derived from ICD-9-CCM administrative data. {\it Medical Care 2002}; {\bf 40}(8): 675-85.


\end{thebibliography}
\end{document}